\documentclass[twocolumn]{aastex63}
\usepackage{graphicx}
\usepackage{natbib}
\usepackage{color}
\usepackage{mathtools}
\usepackage{hyperref} 

\def \bea {\begin{eqnarray}}
\def \ena {\end{eqnarray}}                  
\def \bee {\begin{equation}}
\def \ene {\end{equation}}
\def    \simlt  {\lower.5ex\hbox{$\; \buildrel < \over \sim \;$}}
\def    \simgt  {\lower.5ex\hbox{$\; \buildrel > \over \sim \;$}}

\def    \apjl  		{\rm {ApJL}}

\def	\cm		{\,{\rm {cm}}}
\def	\m		{\,{\rm m}}
\def	\km		{\,{\rm {km}}}
\def	\B		{{\rm B}}
\def	\D		{{\rm D}}
\def	\erg		{\,{\rm {erg}}}
\def	\eV		{\,{\rm {eV}}\,}
\def    \exp 		{\,{\rm {exp}}}
\def	\g		{\,{\rm g}}

\def	\G		{\,{\rm G}}

\def	\K		{\,{\rm K}}

\def	\s		{\,{\rm s}}

\def	\H		{\rm H}

\def	\g		{\,\rm {g}}
\def	\cm		{\,\rm {cm}}

\shorttitle{Evaporative cooling effect}
\shortauthors{Hoang \& Loeb}

\begin{document}
\title{Implications of evaporative cooling by H$_2$ for 1I/`Oumuamua}

\author{Thiem Hoang}
\affiliation{Korea Astronomy and Space Science Institute, Daejeon 34055, Republic of Korea} 
\affiliation{Korea University of Science and Technology, 217 Gajeong-ro, Yuseong-gu, Daejeon, 34113, Republic of Korea}

\author{Abraham Loeb}
\affiliation{Astronomy Department, Harvard University, 60 Garden Street, Cambridge, MA, USA}

\begin{abstract}
The first interstellar object observed in our solar system, 1I/`Oumuamua, exhibited several peculiar properties, including extreme elongation and non-gravitational acceleration. \cite{Bergner.2023} (hereafter BS23) proposed that evaporation of trapped H$_2$ created by cosmic rays (CRs) can explain the non-gravitational acceleration. However, their modeling of the thermal structure of 1I/`Oumuamua ignored the crucial cooling effect of evaporating H$_2$. By taking into account the cooling by H$_2$ evaporation, we show that the surface temperature of H$_2$-water ice is a factor of 9 lower than the case without evaporative cooling. As a result, the thermal speed of outgassing H$_2$ is decreased by a factor of 3. Our one-dimensional thermal modeling that takes into account evaporative cooling for two chosen values of thermal conductivity of $\kappa=0.01$ and $0.1$ WK$^{-1}$m$^{-1}$ shows that the water ice volume available for H$_2$ sublimation at $T>30\K$ would be reduced by a factor of 9 and 5 compared to the results of BS23, not enabling enough hydrogen to propel 1I/`Oumuamua.

\end{abstract}


\section{Introduction}\label{sec:intro}
The detection of the first interstellar object, 1I/`Oumuamua by the Pan-STARRS survey \citep{2017MPEC....U..181B} implies an abundant population of similar interstellar objects (\citealt{Meech:2017hu}; \citealt{2018ApJ...855L..10D}). An elongated shape of semi-axes $\sim 230\m\times 35\m$ is estimated from light-curve modeling (\citealt{2017ApJ...850L..36J}). The extreme axial ratio of $\gtrsim 5:1$ implied by `Oumuamua's lightcurve is mysterious (\citealt{Fraser:2018dg}). 

Several authors \citep{2017ApJ...851L..38B,Gaidos:2017wj} suggested that `Oumuamua is a contact binary, while others speculated that the bizarre shape might be the result of violent processes, such as collisions during planet formation. \cite{Domokos:2017tn} suggested that the elongated shape might arise from ablation induced by interstellar dust, and \cite{Hoang:2018es} suggested that it could originate from rotational disruption of the original body by mechanical torques. \cite{2019Icar..328...14S} suggested that the extreme elongation might arise from planetesimal collisions. The latest proposal involved the tidal disruption of a larger parent object close to a dwarf star \citep{Zhang:2020eu}, but this mechanism is challenged by the preference for a disk-like shape implied by `Oumuamua's lightcurve (\citealt{2019MNRAS.489.3003M}).

The detection of non-gravitational acceleration in the trajectory of `Oumuamua is another peculiarity \citep{Micheli:2018dl}. The authors suggested that cometary activity, such as the outgassing of volatiles, could explain the acceleration excess. Interestingly, no cometary activity of carbon-based molecules was found by deep observations with the Spitzer space telescope \citep{2018AJ....156..261T} and Gemini North telescope \citep{Drahus:2018bd}. \cite{2018ApJ...868L...1B} explained the acceleration anomaly by means of radiation pressure acting on a thin lightsail, and other authors \citep{MoroMartin:2019jf,2020arXiv200810083L,2019arXiv190500935S} suggested a porous object. \cite{Fitzsimmons:2017io} proposed that an icy object of unusual composition might survive its interstellar journey. Previously, \cite{2018A&A...613A..64F} suggested that `Oumuamua might be composed of H$_2$. However, \cite{Rafikov:2018jy} argued that the level of outgassing needed to produce the acceleration excess would rapidly change the rotation period of `Oumuamua, in conflict with the observational data.

Hydrogen ice was suggested by \cite{Seligman:2020vb} to explain `Oumuamua's excess acceleration and unusual shape. Their modeling implied that the object is $\sim$ 100 Myr old. Assuming a speed of $30\km\s^{-1}$, they suggested that the object was produced in a Giant Molecular Cloud (GMC) at a distance of $\sim 5$ kpc. However, their study did not consider the destruction of H$_2$ ice in the interstellar medium (ISM), through evaporation by sunlight. \cite{HoangLoeb:2020b} showed that H$_2$ iceberg could not survive the journey as it would be heated and destroyed by starlight from the GMC birthsite to the solar system. Recent studies by \cite{Jackson.2021,Desch.2021} have proposed that 'Oumuamua is a fragment of N$_2$ ice since
an object of this type is more likely to survive the interstellar journey owing to a much lower sublimation rate. 

\cite{Bergner.2023} (hereafter BS23) proposed that the cosmic-ray (CR) bombardment can dissociate water in the water ice comet and create H$_2$ which are trapped within the CR track under the surface. When 'Oumuamua approaches the Sun, solar radiation heating can cause thermal annealing, which results in the reorganization of the water ice matrix so that H$_2$ entrapped at some pores below the surface can sublimate and escape from the surface, inducing the recoil torque. They found that the temperature of `Oumuamua can reach above $\sim 140$ K at the surface and decreases with the depth for its heliocentric distance below three au (see their figure 3), which is enough for thermal annealing and sublimation of H$_2$. By assuming the well mix of H$_2$ and H$_2$O within the ice, they suggested that the observed acceleration of 1I/`Oumuamua \citep{Micheli:2018dl} can be explained if at least a third of all the water dissociated by CR impact into molecular hydrogen within the iceberg. The temperature profile of the object is a crucial parameter for the release of H$_2$ trapped within the water ice matrix. However, the one-dimensional thermal modeling in \cite{Bergner.2023} ignored the effect of H$_2$ evaporative cooling. Here, we calculate the body temperature profile by taking into account the evaporative cooling and discuss the implications for understanding the true nature of `Oumuamua. 

In Section \ref{sec:surface}, we discuss the heating and cooling mechanisms and calculate the surface temperature and one-dimensional temperature profile of `Oumuamua. Our discussion and conclusions are presented in Section \ref{sec:concl}.

\section{Surface temperature}\label{sec:surface}
In this section, we first consider the heating and cooling processes that occur in the surface layer of 1I/'Oumuamua and derive the surface temperature of the object. Then, we perform detailed calculations for the one-dimensional thermal model of 1I/`Oumuamua.
\subsection{Heating and radiative cooling}
Heating by starlight and solar radiation raises the surface temperature of the H$_2$-water ice. The local interstellar radiation field is assumed to have the same spectrum as the interstellar radiation field (ISRF) in the solar neighborhood \citep{1983A&A...128..212M} with a total radiation energy density of $u_{\rm MMP}\approx 8.64\times 10^{-13} \erg\cm^{-3}$. We normalize the strength of the local radiation field by the dimensionless parameter, $U$, so that the local energy density is $u_{\rm rad}=Uu_{\rm MMP}$. For simplicity, we assume a spherical object shape in our derivations, but the results can be easily generalized to other shapes.

Let $p$ be the albedo of the object surface. The heating rate due to absorption of isotropic interstellar radiation and solar radiation is given by,
\bea
\frac{dE_{\rm abs}}{dt}=\pi R^{2}c\left[Uu_{\rm MMP}+\frac{L_{\odot}}{4\pi cd^{2}}\right](1-p)\epsilon_{\star},\label{eq:dErad}
\ena
where $\epsilon_{\star}$ is the surface emissivity averaged over the background radiation spectrum, and $d$ is the heliocentric distance, i.e., distance from the Sun \citep{HoangLoeb:2020b}. 

In principle, the object can be heated by collisions with ambient gas \citep{HoangLoeb:2020b}, but this process is subdominant in the solar system.

The cooling rate by thermal emission is given by,
\bea
\frac{dE_{\rm emiss}}{dt}=4\pi R^{2}\epsilon_{T}\sigma T^{4},\label{eq:dEemiss}
\ena
where $\epsilon_{T}=\int d\nu \epsilon(\nu)B_{\nu}(T)/\int d\nu B_{\nu}(T)$ is the bolometric emissivity, integrated over all radiation frequencies, $\nu$.

\subsection{Thermal sublimation and evaporative cooling}

The binding energy of H$_2$ to water ice is $E_{b}/k\sim 500\K$ (\citealt{1993ApJ...409L..65S}), equivalent to $E_{b}(\H_{2})\approx 0.05 \eV$. 
H$_2$ can sublimate when the surface temperature is sufficient such that the thermal energy exceeds the binding energy. The characteristic timescale for the evaporation of an H$_2$ molecule from a surface of temperature $T_{\rm surf}$ is
\bea
\tau_{\rm sub}=\nu_{0}^{-1}\exp\left(\frac{E_{b}}{k T_{\rm surf}}\right),\label{eq:tausub}
\ena
where $\nu_{0}$ is the characteristic oscillation frequency of the H$_2$ lattice \citep{1972ApJ...174..321W}. We adopt $\nu_0 = 10^{12} \s^{-1}$ for H$_2$ ice (\citealt{1986ApJ...303...56H}; \citealt{1993ApJ...409L..65S}).

Evaporating H$_2$ molecules carry away heat from the surface and cause evaporative cooling (\citealt{1972ApJ...174..321W}; \citealt{2015ApJ...806..255H}). Let $f({\rm H}_{2}$) be the ratio of H$_2$ to water on the ice surface, i.e., $f({\rm H}_{2})=X_{{\rm H}_{2}:{\rm H}_2{\rm O}}$. Following \cite{HoangLoeb:2020b}, the cooling rate by evaporation of H$_2$ and water is given by,
\bea
\frac{dE_{\rm evap}}{dt}=\frac{E_{b}dN_{\rm mol}}{dt}=\frac{E_{b}N_{s}}{\tau_{\rm sub}(T_{\rm surf})},\label{eq:dEevap}
\ena
where $dN_{\rm mol}/dt$ is the evaporation rate, namely, the number of molecules evaporating per unit time, and $N_{s}=4\pi R^{2}f(H_{2})/r_{s}^{2}$
is the number of surface sites with $r_{s}=10~{\rm \AA}$ being the average size of the H$_2$ surface site, which is taken from that of H$_2$O (see \citealt{1993ApJ...409L..65S}). Here, we neglect the cooling by water because of its much higher sublimation temperature. We also neglect the exothermic effect of annealing/crystallization.

\subsection{Equilibrium surface temperature}

The energy balance between surface heating and cooling is described by
\bea
\frac{dE_{\rm abs}}{dt}=\frac{dE_{\rm emiss}}{dt}+\frac{dE_{\rm evap}}{dt}.\label{eq:balance1}
\ena

Using Equations (\ref{eq:dErad}), (\ref{eq:dEemiss}) and (\ref{eq:dEevap}, one obtains
\bea
\pi R^{2}c\left[Uu_{\rm MMP}+\frac{L_{\odot}}{4\pi cd^{2}}\right](1-p)\epsilon_{\star}&=& 4\pi R^{2}\epsilon_{T}\sigma T^{4}\nonumber\\
&& +\frac{E_{b}N_{s}}{\tau_{\rm sub}(T_{\rm surf})}.\label{eq:balance}
\ena

For our numerical calculations, we adopt the typical albedo value of $p=0.1$ and the interstellar radiation strength of $U=1$. The radius of the object is assumed to be $R=1000$ m, and $\epsilon_{T}=\epsilon_{\star}=1$.

\begin{figure*}
\includegraphics[width = 0.5\textwidth]{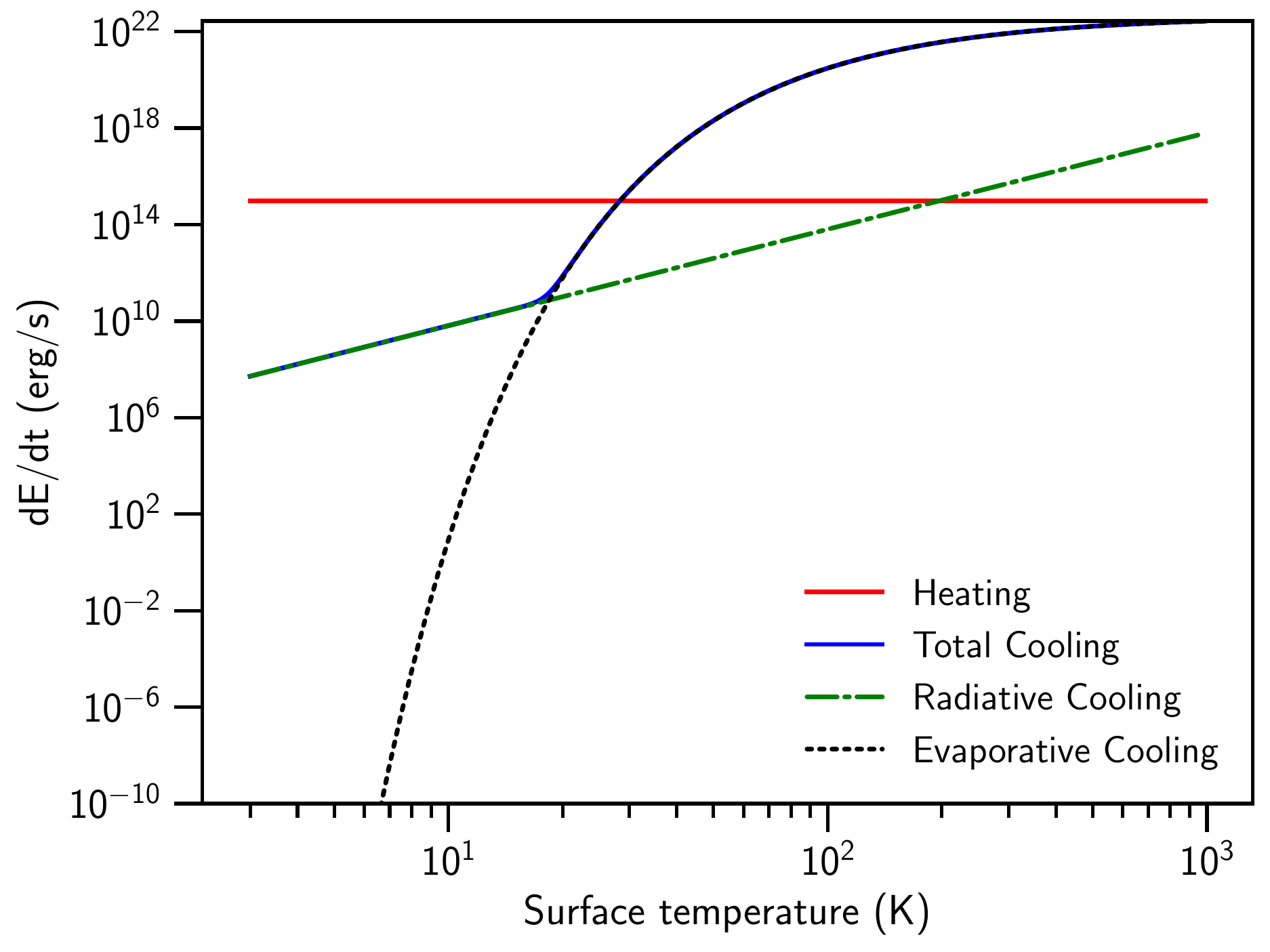}
\includegraphics[width = 0.5\textwidth]{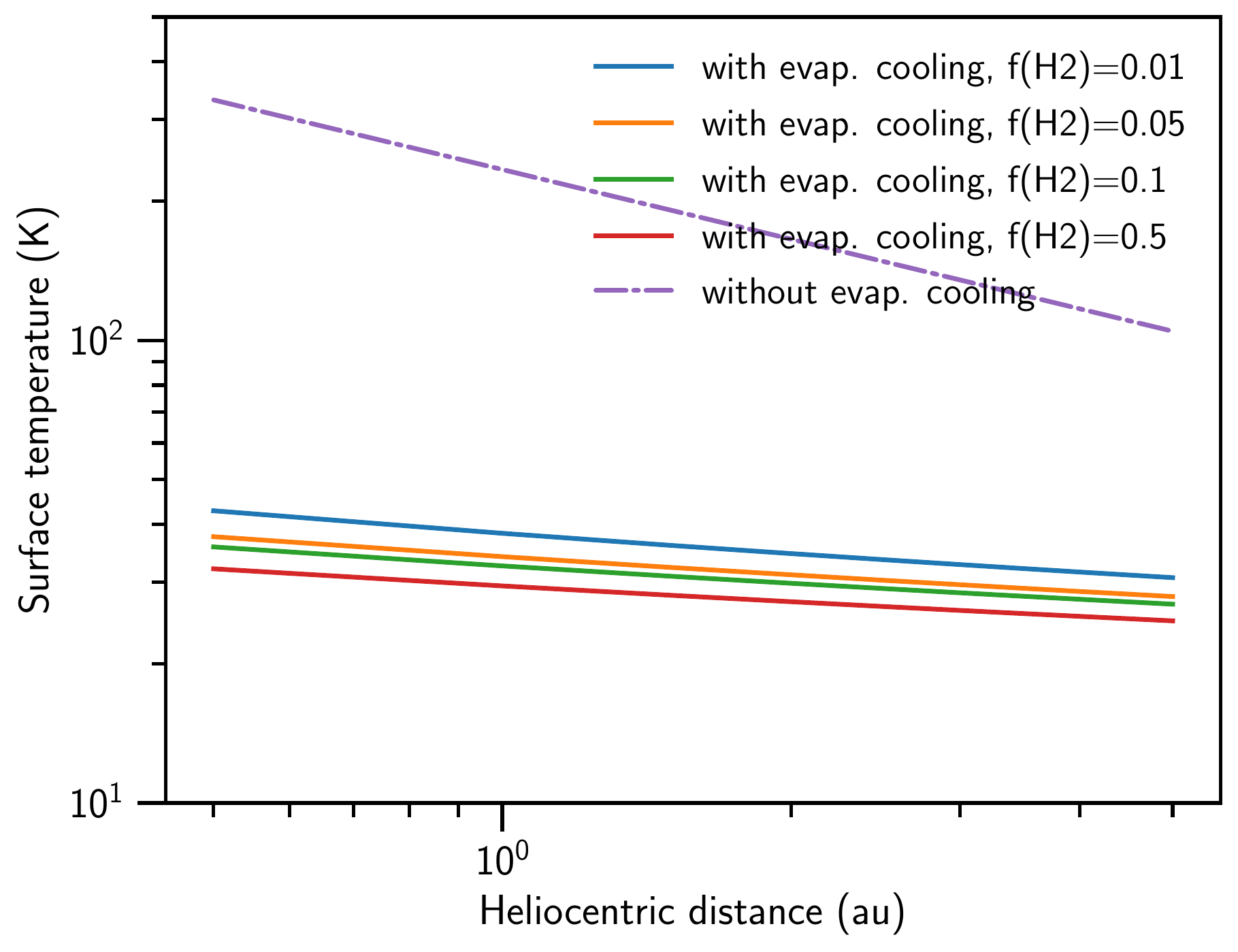}
\caption{Left panel: comparison of heating and cooling rates when the object is located at 1.4 times the Earth separation from the sun. Evaporative cooling by H$_2$ is dominant over radiative cooling. The intersection of heating and total cooling determines the equilibrium surface temperature. Right panel: surface temperature at different distances, calculated for the case with (solid lines) and without (dashed-dotted line) evaporative cooling. Different ratio of H$_2$ to water is assumed. Evaporative cooling by H$_2$ decreases significantly the surface temperature compared to the case without evaporative cooling (dashed-dotted line).}
\label{fig:Tsurf}
\end{figure*}

Figure \ref{fig:Tsurf} (left panel) shows the heating and cooling rates when `Oumuamua is located at 1.4 au from the Sun. Evaporative cooling is dominant over radiative cooling at the temperature above 20 K. Therefore, the effect of evaporative cooling by H$_2$ cannot be ignored as in \cite{Bergner.2023} and must be considered for calculations of the surface temperature.

We numerically solve Equation (\ref{eq:balance}) for the surface equilibrium temperatures for the different distances from the Sun. 
The right panel of Figure \ref{fig:Tsurf} compares the realistic surface temperatures when the evaporating cooling is taken into account with the results without evaporative cooling for the different values of f(H$_2$). For the case without evaporative cooling, our obtained temperature is comparable to those obtained in \cite{Bergner.2023} (see their figure 3). With evaporative cooling, the surface temperature increases slightly with increasing f(H$_2$) for the considered range. As shown, the realistic temperature is significantly lower than the temperature obtained when ignoring the evaporative cooling, by a factor of 9. 

\subsection{One-dimensional thermal model}
Our previous calculations ignore the heat conduction from the surface layer to the inner region and obtain the maximum surface temperature. In this section, we take into account the effect of heat conduction and calculate the body temperature ($T$) as a function of the depth ($z$). 

The heat conduction equation is given by 
\bea
\frac{dT}{dt}=\frac{\kappa}{\rho c_{P}} \frac{d ^{2}T}{dz^{2}},\label{eq:heatcond}
\ena
where $\kappa$ is the thermal conductivity, and $c_{P}$ is the specific heat capacity.

Taking into account the heat conduction and heat diffusivity, Equation \ref{eq:balance}) becomes
\bea
&&\frac{(1-p)}{4}\frac{L_{\odot}}{4\pi d(t)^{2}}= \epsilon_{T} \sigma T_{\rm surf}^{4} + \frac{f(H_{2})E_{b}}{\tau_{\rm sub}(T_{\rm surf})r_{s}^{2}} \nonumber\\
&&+ \left(\kappa \frac{dT}{dz}\right)_{z=0}
+ \rho_{\rm bulk}c_{P}\Delta z \Delta T_{\rm surf}
,\label{eq:surf}
\ena
where the last two terms of the right-hand side describes the heat conduction and heating of the surface layer of the thickness $\Delta z$ and temperature $\Delta T$, respectively. The minor contribution from the ISRF is also ignored. The above equation determines the surface temperature.

We use the numerical code from BS23 that is based on the second Newton-Raphson iterative technique to solve Equation (\ref{eq:heatcond}) for the surface temperature that is constrained by Equation (\ref{eq:surf}). We adopt the same parameters as BS23, with $\rho_{\rm Bulk}=0.5\g\cm^{-3}, \epsilon=0.95, c_{P}=2,000 J~{\rm kg}^{-1}\K^{-1}$. The heat conductivity is chosen for $\kappa=0.01$ and $0.1$ WK$^{-1}$m$^{-1}$.

\begin{figure*}[!htb]
\includegraphics[width = 1.\textwidth]{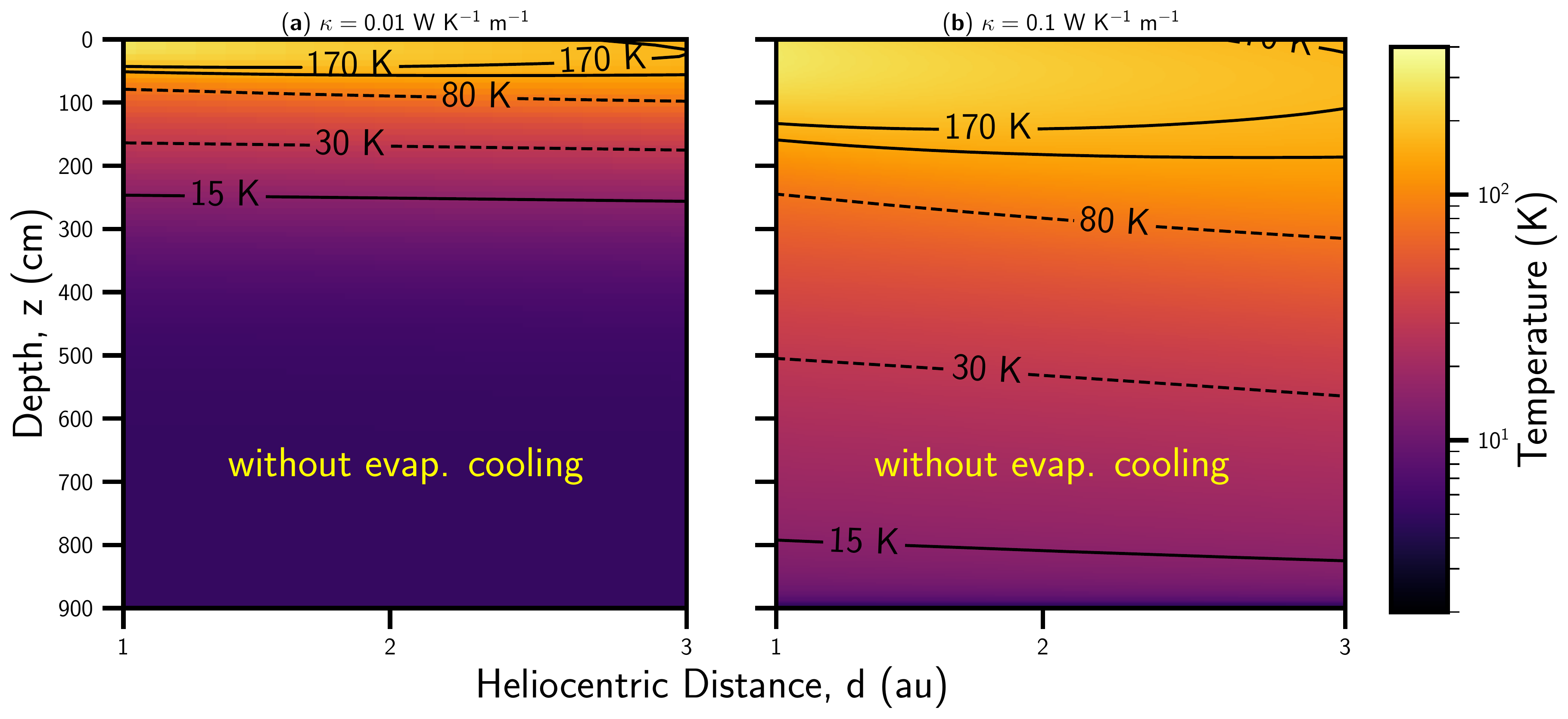}
\includegraphics[width = 1.\textwidth]{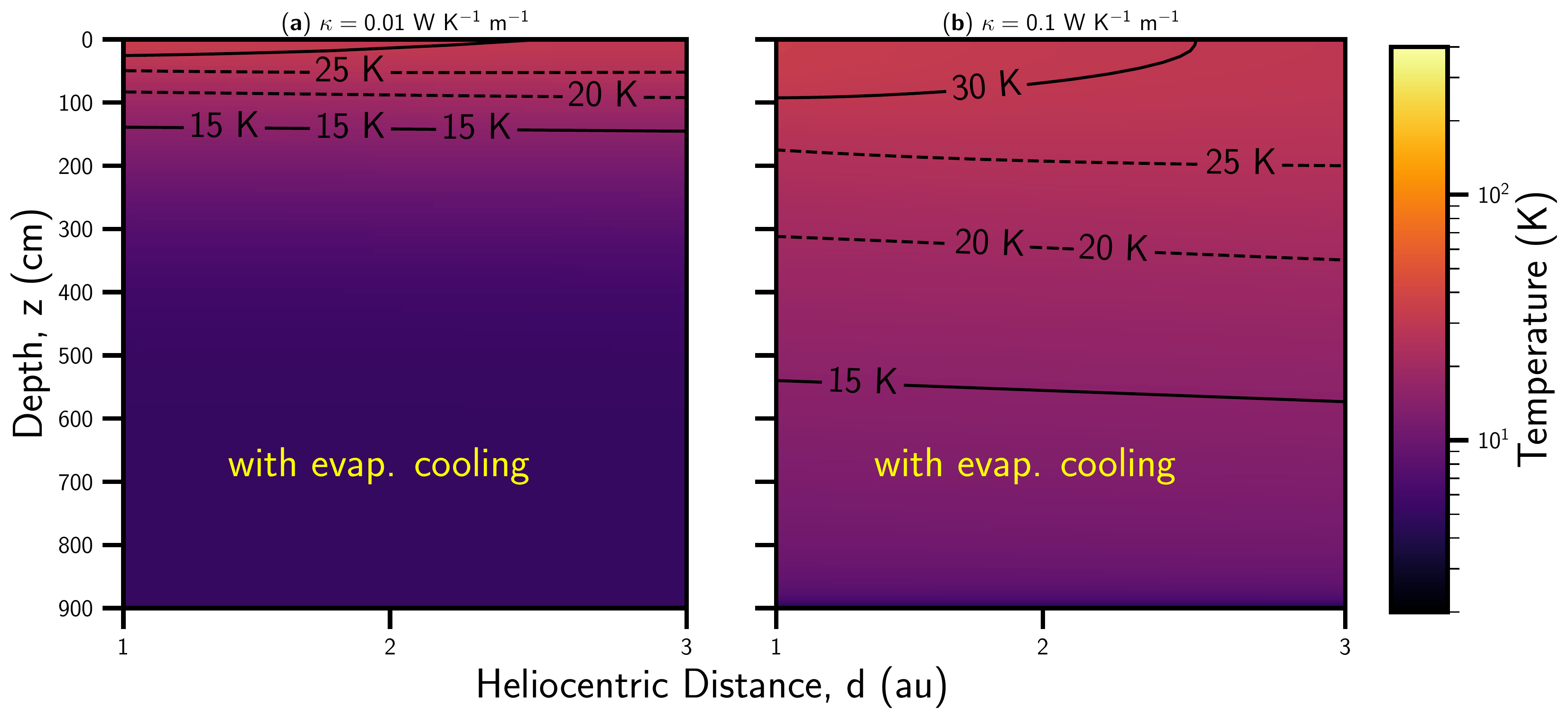}
\caption{Thermal model of 1I/`Oumuamua as a function of the heliocentric distance and depth from the surface for two values of heat conductivity $\kappa$ (left and right panels). Top: Without evaporative cooling. The results are the same as in BS23. Bottom: With evaporative cooling for $f({\rm H}_{2})=0.1$. The temperatures are much lower than the case without evaporative cooling. For larger heat conductivity, heat is rapidly transported, raising the temperature of the body.}
\label{fig:1Dthermal}
\end{figure*}

The top and bottom row of Figure \ref{fig:1Dthermal} shows the thermal model of 1I/'Oumuamua obtained for the case without and with evaporative cooling, respectively. We adopt a value of $f({\rm H}_{2})=0.1$. The body temperature is in general higher for larger conductivity, $\kappa$, due to fast heat conduction. In particular, the body temperature is much lower when evaporative cooling is taken into account. We note that the surface temperature for the case of low conductivity of $\kappa=0.01$ WK$^{-1}$m$^{-1}$ is similar to the result obtained in Figure \ref{fig:Tsurf}.

From Figure \ref{fig:1Dthermal}, one can see that, for the lower conductivity, the critical depth at which the temperature is still high enough for H$_2$ sublimation, take at 15 K (see BS23), $z_{T=15\K}$ decreases by a factor of 2 (see panels (a)). For the higher conductivity (see panels (b)), the depth of $z_{T=15\K}$ decreases by a factor of 1.5. For thermal annealing which requires the minimum temperature of 30 K, the critical depth is more significantly reduced, by a factor of $z_{T=30\K}^{\rm evap}/z_{T=30\K}^{\rm noevap}\sim 9$ and $\sim 5$.

\section{Discussion and conclusions}\label{sec:concl}
More than 5 years after the discovery of 1I/`Oumuamua, many peculiar properties of it are still hotly debated. \cite{Bergner.2023} proposed that evaporation of trapped H$_2$ created by cosmic rays (CRs) by sunlight can explain the non-gravitational acceleration. However, their calculations of surface and body temperature disregard the cooling effect by evaporating H$_2$. 

To quantify the effect of evaporative cooling, we first calculated the surface temperature in the presence of evaporative cooling using an energy conservation equation without heat conduction. We found that the resulting surface temperatures are much lower than the case without evaporative cooling (see Figure \ref{fig:Tsurf}, right panel). For instance, at a distance of 1 au, our surface temperature is a factor of 9 lower than the case without evaporative cooling. As a result, the evaporating cooling rate is then larger by a factor of $e^{9}=10^{4}$, and the thermal speed of outgassing H$_2$ is consequently decreased by a factor of $\sqrt{9}=3$.

We also constructed a one-dimensional thermal model of `Oumuamua by taking into account the evaporative cooling by H$_2$ evaporation. Our results in Figure \ref{fig:1Dthermal} show the subsurface temperature decreases to below 15-30 K at the depth below 10 cm from the surface. Therefore, the thermal annealing of amorphous water ice, a key process that is appealed to by \cite{Bergner.2023} to release H$_2$, only occurs in the thin warm surface layer with temperature exceeding the annealing threshold of $T_{\rm anneal}\sim 30$ K. Therefore, the amount of H$_2$ available for outgassing is reduced by a factor of $(z_{T=15\K}^{\rm evap}/z_{T=15\K}^{\rm noevap})\sim 1.5$ and $2$ for two chosen values of $\kappa$. In particular, the amount of water ice that can undergo thermal annealing at $T>30$ K is significantly reduced by a factor of $(z_{T=30\K}^{\rm evap}/z_{T=30\K}^{\rm noevap})\sim 9$ and $5$ for the two chosen values of $\kappa$.

The reduction in the water ice volume suitable for thermal annealing at $T \gtrsim 30$ K by a factor of 9 and 5 compared to the results of BS23, makes the model untenable since not enough hydrogen will be available to propel `Oumuamua. Even if a thin surface layer happens to be made of pure molecular hydrogen, it will not survive the journey through interstellar space as a result of heating by starlight \citep{HoangLoeb:2020b}. Long-period comets from the Oort cloud, which originate outside the heliosphere, are exposed to the same interstellar conditions as interstellar comets and do not show pure hydrogen propulsion. Indeed, the truly interstellar comet, 2I/Borisov, resembled solar system comets~\citep{2021A&A...650L..19O}.


Finally, it is noted that our calculations assumed that the sublimation of H$_2$ occurs from the surface layer. Note that, when modeling outgassing and NGA, BS23 also assumed a well-mixed H:H$_2$O throughout the body. In realistic situations, sublimation can occur within the object's volume and would even consume more energy. However, sublimating H$_2$ must escape into the space in order to produce recoil and explain the non-gravitational acceleration. Therefore, sub-surface sublimation eventually results in the loss of surface energy and causes the evaporative cooling. Moreover, given its long journey in the ISM, the probability that all H$_2$ entrapped below the surface of `Oumuamua and undergo the sub-surface sublimation is unlikely. Therefore, our results of the body temperature profile correspond to a conservative limit of the body temperature. A detailed modeling of the location of sublimation and energy transfer of sublimating H$_2$ to the body before reaching the surface is required for accurately understanding the role of H$_2$ on the excess acceleration of 'Oumuamua.

\acknowledgements
We thank the referee for comments that helped improve the paper.
T.H. acknowledges the support by the National Research Foundation of Korea (NRF) grant funded by the Korea government (MSIT) (2019R1A2C1087045). AL was supported in part by the Galileo Project and the Black Hole Initiative at Harvard University. This work was partly supported by a grant from the Simons Foundation to IFIRSE, ICISE (916424, N.H.).

\bibliography{ms.bbl}

\begin{thebibliography}{}
\expandafter\ifx\csname natexlab\endcsname\relax\def\natexlab#1{#1}\fi

\bibitem[{{Bacci} {et~al.}(2017){Bacci}, {Maestripieri}, {Tesi}, \&
  et~al.}]{2017MPEC....U..181B}
{Bacci}, P., {Maestripieri}, M., {Tesi}, L., \& et~al. 2017, MPEC, U181

\bibitem[{Bannister {et~al.}(2017)Bannister, Schwamb, Fraser, Marsset,
  Fitzsimmons, Benecchi, Lacerda, Pike, Kavelaars, Smith, Stewart, Wang, \&
  Lehner}]{2017ApJ...851L..38B}
Bannister, M.~T., Schwamb, M.~E., Fraser, W.~C., {et~al.} 2017, \apjl, 851, L38

\bibitem[{{Bergner} \& {Seligman}(2023)}]{Bergner.2023}
{Bergner}, J.~B., \& {Seligman}, D.~Z. 2023, \nat, 615, 610

\bibitem[{Bialy \& Loeb(2018)}]{2018ApJ...868L...1B}
Bialy, S., \& Loeb, A. 2018, \apjl, 868, L1

\bibitem[{Desch \& Jackson(2021)}]{Desch.2021}
Desch, S.~J., \& Jackson, A.~P. 2021, Journal of Geophysical Research: Planets,
  doi:10.1029/2020je006807

\bibitem[{Do {et~al.}(2018)Do, Tucker, \& Tonry}]{2018ApJ...855L..10D}
Do, A., Tucker, M.~A., \& Tonry, J. 2018, \apjl, 855, L10

\bibitem[{Domokos {et~al.}(2017)Domokos et al.}]{Domokos:2017tn}
Domokos, G., Sipos, A. {\'A}., Szab{\'o}, G.~M., et al.\ 2017, RNAAS, 1, 50

\bibitem[{Drahus {et~al.}(2018)Drahus, Guzik, Waniak, Handzlik, Kurowski, \&
  Xu}]{Drahus:2018bd}
Drahus, M., Guzik, P., Waniak, W., {et~al.} 2018, \natas, 1

\bibitem[{Fitzsimmons {et~al.}(2018)Fitzsimmons, Snodgrass, Rozitis, Yang,
  Hyland, Seccull, Bannister, Fraser, Jedicke, \& Lacerda}]{Fitzsimmons:2017io}
Fitzsimmons, A., Snodgrass, C., Rozitis, B., {et~al.} 2018, \natas, 2, 133

\bibitem[{{Fraser} {et~al.}(2018){Fraser}, {Pravec}, {Fitzsimmons}, {Lacerda},
  {Bannister}, {Snodgrass}, \& {Smoli{\'c}}}]{Fraser:2018dg}
{Fraser}, W.~C., {Pravec}, P., {Fitzsimmons}, A., {et~al.} 2018, \natas, 2, 383

\bibitem[{F{\"u}glistaler \& Pfenniger(2018)}]{2018A&A...613A..64F}
F{\"u}glistaler, A., \& Pfenniger, D. 2018, \aa, 613, A64

\bibitem[{Gaidos(2018)}]{Gaidos:2017wj}
Gaidos, E. 2018, \mnras, 477, 5692

\bibitem[{Hegyi \& Olive(1986)}]{1986ApJ...303...56H}
Hegyi, D.~J., \& Olive, K.~A. 1986, \apj, 303, 56

\bibitem[{Hoang {et~al.}(2015)Hoang, Lazarian, \&
  Schlickeiser}]{2015ApJ...806..255H}
Hoang, T., Lazarian, A., \& Schlickeiser, R. 2015, \apj, 806, 255

\bibitem[{{Hoang} \& {Loeb}(2020)}]{HoangLoeb:2020b}
{Hoang}, T., \& {Loeb}, A. 2020, \apjl, 899, L23

\bibitem[{Hoang {et~al.}(2018)Hoang, Loeb, Lazarian, \& Cho}]{Hoang:2018es}
Hoang, T., Loeb, A., Lazarian, A., \& Cho, J. 2018, \apj, 860, 42

\bibitem[{Jackson \& Desch(2021)}]{Jackson.2021}
Jackson, A.~P., \& Desch, S.~J. 2021, Journal of Geophysical Research: Planets,
  doi:10.1029/2020je006706

\bibitem[{Jewitt {et~al.}(2017)Jewitt, Luu, Rajagopal, Kotulla, Ridgway, Liu,
  \& Augusteijn}]{2017ApJ...850L..36J}
Jewitt, D., Luu, J., Rajagopal, J., {et~al.} 2017, \apjl, 850, L36

\bibitem[{Luu {et~al.}(2020)Luu, Flekk{\o}y, \&
  Toussaint}]{2020arXiv200810083L}
Luu, J.~X., Flekk{\o}y, E.~G., \& Toussaint, R. 2020, \apjl, 900, L22

\bibitem[{Mashchenko(2019)}]{2019MNRAS.489.3003M}
Mashchenko, S. 2019, \mnras, 489, 3003

\bibitem[{Mathis {et~al.}(1983)Mathis, Mezger, \&
  Panagia}]{1983A&A...128..212M}
Mathis, J.~S., Mezger, P.~G., \& Panagia, N. 1983, \aa, 128, 212

\bibitem[{Meech {et~al.}(2017)Meech, Weryk, \& Micheli}]{Meech:2017hu}
Meech, K.~J., Weryk, R., \& Micheli, M. e.~a. 2017, Nature, 552, 378

\bibitem[{{Micheli} {et~al.}(2018){Micheli}, {Farnocchia}, {Meech}, {Buie},
  {Hainaut}, {Prialnik}, {Sch{\"o}rghofer}, {Weaver}, {Chodas}, {Kleyna},
  {Weryk}, {Wainscoat}, {Ebeling}, {Keane}, {Chambers}, {Koschny}, \&
  {Petropoulos}}]{Micheli:2018dl}
{Micheli}, M., {Farnocchia}, D., {Meech}, K.~J., {et~al.} 2018, \nat, 559, 223

\bibitem[{Moro-Martin(2019)}]{MoroMartin:2019jf}
Moro-Martin, A. 2019, \apjl, 872, 0

\bibitem[{{Opitom} {et~al.}(2021){Opitom}, {Jehin}, {Hutsem{\'e}kers},
  {Shinnaka}, {Manfroid}, {Rousselot}, {Raghuram}, {Kawakita}, {Fitzsimmons},
  {Meech}, {Micheli}, {Snodgrass}, {Yang}, \& {Hainaut}}]{2021A&A...650L..19O}
{Opitom}, C., {Jehin}, E., {Hutsem{\'e}kers}, D., {et~al.} 2021, \aap, 650, L19

\bibitem[{Rafikov(2018)}]{Rafikov:2018jy}
Rafikov, R.~R. 2018, The Astrophysical Journal Letters, 867, 0

\bibitem[{Sandford \& Allamandola(1993)}]{1993ApJ...409L..65S}
Sandford, S.~A., \& Allamandola, L.~J. 1993, \apj, 409, L65

\bibitem[{Sekanina(2019)}]{2019arXiv190500935S}
Sekanina, Z. 2019, arXiv, 1905.00935v3

\bibitem[{Seligman \& Laughlin(2020)}]{Seligman:2020vb}
Seligman, D., \& Laughlin, G. 2020, \apjl, 896, L8

\bibitem[{Sugiura {et~al.}(2019)Sugiura, Kobayashi, \&
  Inutsuka}]{2019Icar..328...14S}
Sugiura, K., Kobayashi, H., \& Inutsuka, S.-i. 2019, Icarus, 328, 14

\bibitem[{Trilling {et~al.}(2018)Trilling, Mommert, Hora, Farnocchia, Chodas,
  Giorgini, Smith, Carey, Lisse, Werner, McNeill, Chesley, Emery, Fazio,
  Fernandez, Harris, Marengo, Mueller, Roegge, Smith, Weaver, Meech, \&
  Micheli}]{2018AJ....156..261T}
Trilling, D.~E., Mommert, M., Hora, J.~L., {et~al.} 2018, \aj, 156, 261

\bibitem[{Watson \& Salpeter(1972)}]{1972ApJ...174..321W}
Watson, W.~D., \& Salpeter, E.~E. 1972, \apj, 174, 321

\bibitem[{Zhang \& Lin(2020)}]{Zhang:2020eu}
Zhang, Y., \& Lin, D.~N. 2020, \natas, 4, 852

\end{thebibliography}

\end{document}